# Millimeter Wave Channel Modeling via Generative Neural Networks

William Xia[†]    Sundeep Rangan[†]    Marco Mezzavilla[†]    Angel Lozano[♭]
Giovanni Geraci[♭]    Vasilii Semkin[♯]    Giuseppe Loianno[†]
[†]NYU Tandon School of Engineering, Brooklyn, NY, USA
[♯]VTT Technical Research Centre of Finland Ltd, Finland
[♭]Univ. Pompeu Fabra, Barcelona, Spain

*Abstract*—Statistical channel models are instrumental to design and evaluate wireless communication systems. In the millimeter wave bands, such models become acutely challenging; they must capture the delay, directions, and path gains, for each link and with high resolution. This paper presents a general modeling methodology based on training generative neural networks from data. The proposed generative model consists of a two-stage structure that first predicts the state of each link (line-of-sight, non-line-of-sight, or outage), and subsequently feeds this state into a conditional variational autoencoder that generates the path losses, delays, and angles of arrival and departure for all its propagation paths. Importantly, minimal prior assumptions are made, enabling the model to capture complex relationships within the data. The methodology is demonstrated for $28\,\text{GHz}$ air-to-ground channels in an urban environment, with training datasets produced by means of ray tracing.

## I. Introduction

The design and evaluation of any wireless communication system hinges critically on the availability of statistical channel models that adequately describe the distribution of constituent parameters in the scenarios of interest. Indeed, statistical models have been the foundation of virtually every cellular and WLAN commercial evaluation methodology for decades. The extension of these models to the millimeter wave (mmWave) bands, however, is challenging [1]: systems operate over broad bandwidths and with highly directive antenna arrays, and thus require models that capture the delay, directions, and path gains, with sufficient resolution to properly evaluate beamforming, equalization, and other key algorithms [2], [3]. The parameters in these models can exhibit utterly complex relationships that are very difficult to establish from first principles.

In such a context, data-driven machine-learning methods are an attractive recourse that entails minimal assumptions and can naturally capture intricate probabilistic relationships. Neural networks (NNs) have been specifically advocated in [4]–[7] for mmWave channel modeling, whereby upon an input corresponding to some location within an indoor system, the NN outputs the model parameters for that location; in essence, the parameters are then a regression from the training dataset, much as in data-based signal power maps and in learning-based planning and prediction tools [8], [9]. A strong aspect of all these works is their inherent site-specific nature, a virtue when it comes to optimizing specific deployments. Alternatively, there is interest in models that can produce channel parameters broadly representative of some general environment, say an urban microcell system.

Generative NNs, which have proven enormously successful with images and text [10]–[12], offer a natural approach to data-driven channel modeling that can broadly represent complex settings, and some early works have successfully trialed generative adversarial networks (GANs) for simple wireless channels [13], [14]. The present paper propounds a powerful and widely applicable methodology for generative NN channel modeling. For data provisioning, we rely on ray tracing (specifically the tool [15]), which has developed substantially for mmWave communication [16], [17] and can supply the large datasets needed to train large NNs. The proposed methodology has the following attributes:

- The wideband, double-directional nature of the channel is captured, meaning the delay, path loss, and angular information on all paths for each link. This description is compatible with 3GPP evaluation methodologies [1], [18] and can provide the full wideband MIMO response given specific antenna configurations at transmitter and receiver. No prior assumptions are made regarding the relations between parameters, and the model is able to capture complex and interesting data relationships.
- The generative model features a novel two-stage structure where a first NN determines if the link is line-of-sight (LOS), non-line-of-sight (NLOS), or in outage, with a second stage that employs a conditional variational autoencoder (VAE) to predict the link parameters.

The methodology is demonstrated by characterizing $28\,\text{GHz}$ channels connecting unmanned aerial vehicles (UAVs) with both street-level and rooftop-mounted receivers. This use case is of great interest, as the latest standard-defined air-to-ground model is only calibrated at sub-6 GHz frequencies [18]. Channels for aerial communication also present unique challenges such as the parameter dependencies on the unmanned aerial

S. Rangan, W. Xia, and M. Mezzavilla were supported by NSF grants 1302336, 1564142, 1547332, and 1824434, NIST, SRC, and the industrial affiliates of NYU WIRELESS. A. Lozano and G. Geraci were supported by the ERC grant 694974, by MINECO's Project RTI2018-101040, and by the Junior Leader Fellowship Program from "la Caixa" Banking Foundation.

vehicle (UAV) altitude, their 3D orientation, or the building heights [1], [18]–[21]. For example, [22] proposes an empirical propagation model for UAV-to-UAV communication at 60 GHz, which applies to altitude values between 6 and 15 m. However, the aerial measurement campaign does not include NLOS links and hence does not characterize reflections and diffraction in the face of blockage.

The generative NN model developed in this work is publicly available [23].

## II. PROBLEM FORMULATION

We consider the modeling of channels linking a transmitter with a receiver. In the aerial context, we take the UAV to be the transmitter and the base station (or gNB in 3GPP terminology [18]) to be the receiver, yet, owing to reciprocity, the roles of transmitter and receiver are interchangeable. Each link is described by the set of parameters [24]

$$\mathbf{x} = \left\{ \left( L_k, \phi_k^{\text{rx}}, \theta_k^{\text{rx}}, \phi_k^{\text{tx}}, \theta_k^{\text{tx}}, \tau_k \right), \ k = 1, \ldots, K \right\}, \quad (1)$$

where $K$ is the number of paths and, for each path $k$, $L_k$ is the pathloss, $(\phi_k^{\text{rx}}, \theta_k^{\text{rx}})$ are its azimuth and elevation angles of arrival, $(\phi_k^{\text{tx}}, \theta_k^{\text{tx}})$ are its azimuth and elevation angles of departure, and $\tau_k$ is the absolute propagation delay. Unlike standard 3GPP spatial cluster models (e.g. [1]), we do not consider angular or delay dispersion within each path. This is not a limitation of the model, though, but only a consequence of the tool that produces the training data not accommodating diffuse reflections. If angular or delay spread data were available, these aspects could be modeled as well.

For streamlining purposes, the number of paths in the model is fixed to some value $K = K_{\max}$ with $L_k = L_{\max}$ for paths that are not actually present. We set $K_{\max} = 20$ paths and $L_{\max} = 200$ dB, which is compatible with the maximum pathloss detectable by the ray tracer. With these settings, the data vector $\mathbf{x}$ in (1) contains $6K_{\max} = 120$ parameters per link.

Let

$$\mathbf{u} = (\mathbf{d}, c) \quad (2)$$

denote the *link condition* vector, with $\mathbf{d} = (d_x, d_y, d_z)$ the vector connecting the UAV with the gNB and with $c$ the type of gNB. As described in Section IV, for the UAV application, two types of gNBs are considered: terrestrial street-level gNBs and aerial roof-mounted gNBs. The goal is to capture the conditional distribution $p(\mathbf{x}|\mathbf{u})$ over some ensemble of possible links. That is, we wish to model the distribution of the paths in a link as a function of the link conditions for some environment. As anticipated in the introduction, we consider a generative scheme whereby we model $\mathbf{x}$ as

$$\mathbf{x} = g(\mathbf{u}, \mathbf{z}), \quad (3)$$

where $\mathbf{z}$ is a random vector, termed the *latent vector*, with some fixed prior distribution, $p(\mathbf{z})$ while $g(\mathbf{u}, \mathbf{z})$ is the *generating function*, to be trained from data.

TABLE I: Generative model parameters

| Item | Value | | |
|---|---|---|---|
| | Link state prediction | Path VAE encoder | Path VAE decoder |
| Number of inputs | 5 | 5 + 120 | 5 + 20 |
| Hidden units | [25, 10] | [200, 80] | [80, 200] |
| Number of outputs | 3 | 20 + 20 | 120 + 120 |
| Optimizer | Adam | Adam | |
| Learning rate | $10^{-3}$ | $10^{-4}$ | |
| Epochs | 50 | 10000 | |
| Batch size | 100 | 100 | |
| Number of parameters | 1653 | 44520 | 40720 |

Once trained, generative models are conveniently applicable in simulations: the locations of UAVs and gNBs are typically generated stochastically according to some deployment model, providing the condition vector $\mathbf{u}$ for each link. Random vectors $\mathbf{z}$ can then be produced for each link from the prior $p(\mathbf{z})$ and, from $\mathbf{u}$ and $\mathbf{z}$, the path parameters $\mathbf{x} = g(\mathbf{u}, \mathbf{z})$ follow. These parameters can be generated for both intended and interfering links and, in conjunction with the antenna patterns, array configuration, and beam tracking methods, they allow computing quantities of interest such as the SNRs or bit rates.

## III. PROPOSED GENERATIVE MODEL

The propounded generative model, depicted in Fig. 1, consists of two cascaded NNs, namely (a) a link-state prediction network, followed by (b) a path generative network.

### A. Link-State Predictor

As recognized by standard 3GPP models such as [1], it is crucial to first determine the existence or lack of the LOS path. To this end, the link-state-prediction NN accepts the condition $\mathbf{u}$ in (2) and produces probabilities for the link being in one of three states [25]:

1) LOS: The LOS path is present, possibly in addition to non-LOS (NLOS) paths;
2) NLOS: The LOS path is absent, but at least one NLOS path is active; and
3) NoLink: No propagation paths (either LOS or NLOS) exist for this link.

To model the link state probabilities, we employ a fully connected NN configured as per Table I. The input to this NN is the condition vector, $\mathbf{u} = (d_x, d_y, d_z, c)$. A fixed nonlinear transformation is applied to map this to a 5-dimensional feature space separating the horizontal and vertical distance and gNB types. The five-dimensional transformed input is is then passed through a standard scaler, and subsequently through two hidden layers. The output is a three-way softmax corresponding to the three states. The link state is then sampled from the probabilities from the softmax output. In the sequel, we let $s \in \{\texttt{LOS}, \texttt{NLOS}, \texttt{NoLink}\}$ denote the link state output.

### B. Path Generator

The goal of the path generator is to create the parameters in (1) for the NLOS paths, given the condition $\mathbf{u}$ and link state $s$.

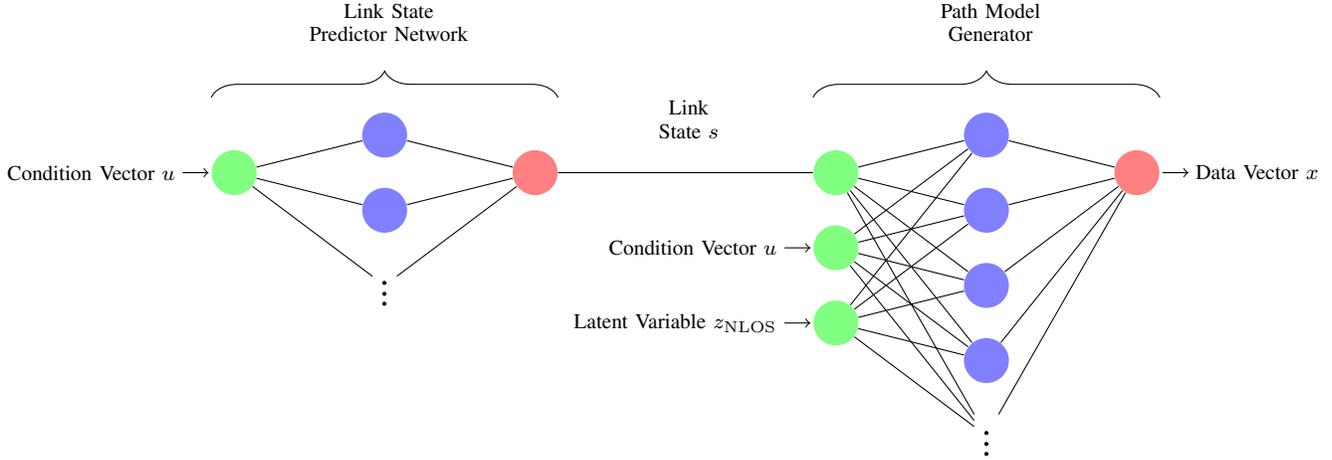

Fig. 1: Overall architecture for the two-stage generative model.

For the LOS path, when it exists, the delay and angles of departure and arrival can be computed deterministically from the geometry while its path loss can be computed from Friis' law [24].

Let $x_{NLOS}$ denote the NLOS components of the path vector $x$ in (1) and let $z_{NLOS}$ denote the corresponding component of $z$. The path generator is thus a function, $x_{NLOS} = g_{NLOS}(u, s, z_{NLOS})$, that takes the link condition $u$, the link state $s$, and $z_{NLOS}$, to generate $x_{NLOS}$. Ideally, the path generator should be trained so that the conditional distribution of $x_{NLOS}$ given $u, s$ matches the conditional distribution found in the data. There are a large number of methods for training generative models, the two most common being variants of GANs [10], [11] or VAEs [12]. We found the most success with a VAE, as it avoids the minimax optimization required by a GAN.

In the VAE paradigm, the generator $x_{NLOS} = g_{NLOS}(u, s, z_{NLOS})$ is termed the *decoder*. The VAE also requires training a so-called *encoder* that maps data samples $x_{NLOS}$ and $u, s$, back to the latent variables $z_{NLOS}$. This encoder attempts to approximate sampling from the posterior density of $z_{NLOS}$ given $(x_{NLOS}, u, s)$. The encoder and decoder are then jointly optimized to maximize an approximation of the log-likelihood called evidence lower bound (ELBO); see [12] for details.

In our case, the encoder and decoder are embodied by fully connected NNs configured as per Table I. The latent variable is realized as a 20-dimensional Gaussian vector. The decoder accepts the 20-dimensional Gaussian vector plus 5 transformed conditioned and link state variables, outputting means and variances on the 120-dimensional vector $x_{NLOS}$ for a total of $120 + 120$ outputs. Similarly, the encoder network takes 5 conditioned variables and a 120-dimensional data input and produces means and variances for the 20-dimensional random latent variable.

## IV. AIR-TO-GROUND RAY TRACING DATA AT 28 GHZ

Experimental data on UAV channels has been limited, particularly in the mmWave bands [19], [20], [26]–[28]. In this work, we employ a powerful ray tracing package, Wireless InSite by Remcom [15], which was also used in [29]. A 3D representation of a region measuring 500 m × 500 m and corresponding to Reston, VA was imported. The representation, shown in Fig. 2, includes terrain and building data. Receiving gNBs were manually placed at 120 locations:

- *80 terrestrial gNBs:* These sites were placed on streets approximately 2 m high, emulating typical locations for current 5G picocells designed to serve ground users. We are interested in these locations for aerial channel modeling, both to see whether terrestrial cells can serve UAVs and to understand the interference between UAV and terrestrial communication.
- *40 aerial rooftop gNBs:* These sites were located on rooftops, typically 30 m above street level. Such sites could be used for providing coverage to UAVs, particularly at high altitudes.

Transmitting UAVs were placed at 180 locations in the 3D volume. Specifically, the UAVs were placed at 60 different $(x, y)$ locations in the area with three different altitudes in each point. This creates a total of $180 \times 120 = 21600$ links, i.e., UAV-gNB pairs. The Wireless InSite tool was then run to simulate the channel for each link. The output of the tool produces the path data $x$ in (1). Although not used here, the ray tracing also produces the full route of each path including the scattering locations. All simulations were conducted at 28 GHz, the dominant carrier frequency for emerging 5G mmWave systems.

For ray tracing purposes, the maximum number of reflections is set to 6 and the maximum number of diffractions is set to 1. The material is set to concrete with a permittivity of 5.31 F/m for both the ground and wall surfaces. As its output, the simulator provides the directions of arrival, directions of departure, and path losses for each link.

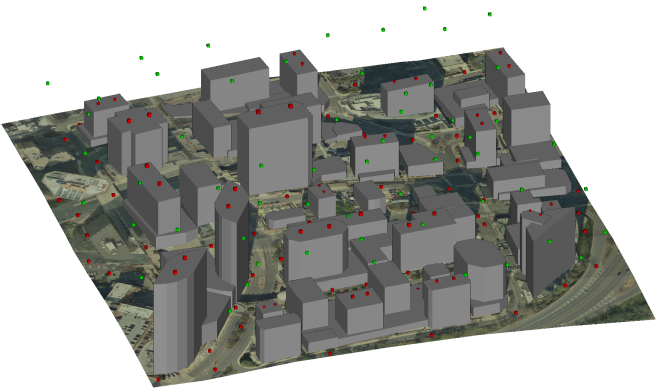

Fig. 2: Ray tracing simulation area representing a 500m × 500m region of Reston, VA. Shown are 60 of the 180 UAV locations (green dots), as well as the terrestrial and rooftop aerial gNB locations (red dots).

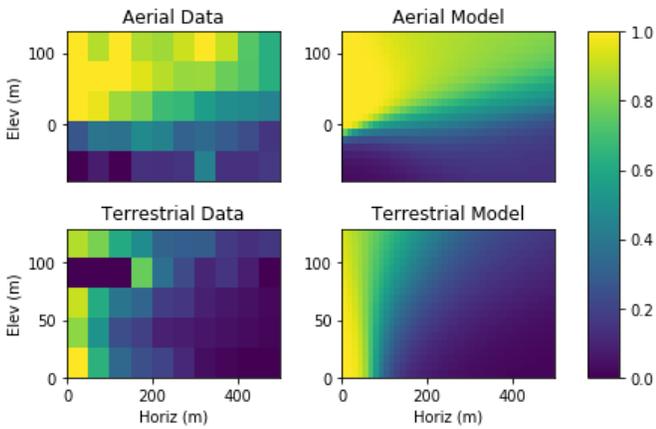

Fig. 3: Conditional probability of a LOS link as a function of horizontal and vertical position relative to the base station for aerial and terrestrial types. Left: Empirical distribution on the test data; Right: Probability from the trained link-state predictor.

## V. UAV MMWAVE MODELING RESULTS

The data set of 21600 links was divided into 70% for training and 30% for test. All code was implemented in Tensorflow 2.2 and the code, data and pre-trained models can be found in [23]. This section describes various features of the learned model and its ability to capture interesting wireless phenomena.

### A. LOS Probability

First, to illustrate the functioning of the link state predictor, Fig. 3 plots the conditional probability of the link being in the LOS state as a function of its horizontal and vertical distances. The probability is separately plotted for aerial gNBs (top plots) and terrestrial gNBs (bottom plots). Similar plots could also be produced for the `NLOS` and `NoLink` states. The left-hand-side plot shows the empirical probability as measured on the test data and the right-hand-side plot shows the probability from the output of the trained link-state predictor.

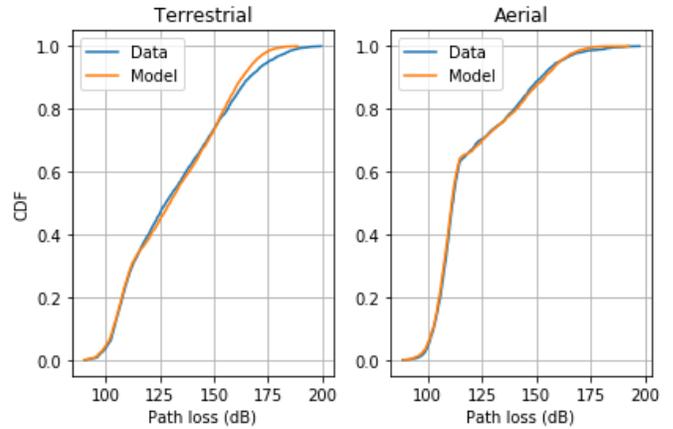

Fig. 4: CDF of the path loss for the links in LOS or NLOS states with the distribution of the positions taken from the test data.

The link-state predictor matches the basic trends of the empirical distribution, and it reflects the very different behavior between terrestrial and aerial gNBs. In particular, the aerial gNBs can provide high probabilities of LOS coverage at long horizontal distances provided the UAV is high enough. In contrast, terrestrial gNBs are much more limited in terms of horizontal coverage.

### B. Path Loss with Omnidirectional Antennas

We now turn to evaluating the accuracy of the rest of the model parameters. Fundamentally, we want to measure how close the distribution of the trained generative model $\mathbf{x} = g(\mathbf{u}, \mathbf{z})$ in (3) is to the observed conditional distribution of the test data. To this end, let $(\mathbf{u}_i, \mathbf{x}_i)$, $i = 1, \ldots, N_{\text{ts}}$ be the set of test data where each sample contains a link condition $\mathbf{u}_i$ and its corresponding path data vector $\mathbf{x}_i$. To evaluate how closely the learned model fits this test data, for each sample we can compute some statistic $v_i = \phi(\mathbf{u}_i, \mathbf{x}_i)$. The statistic should be of some relevance to the application. As an example, we compute the path loss that would be experienced if UAV and gNB were equipped with omnidirectional antennas.

Using the same conditions $\mathbf{u}_i$ from the test data, we generate a random sample $\mathbf{x}_i^{\text{rnd}} = g(\mathbf{u}_i, \mathbf{z}_i)$ from the trained model $g(\mathbf{u}, \mathbf{z})$ and a random $\mathbf{z}_i$. We can then compute the statistics $v_i^{\text{rnd}} = \phi(\mathbf{u}_i, \mathbf{x}_i^{\text{rnd}})$ and compare the CDFs of $v_i^{\text{rnd}}$ and $v_i$.

Fig. 4 shows the empirical CDFs of path loss for the test data and for the model, with the same condition values. An excellent match is observed for both aerial and terrestrial gNBs. In particular, the trained generative model is able to capture the dual-slope nature of the CDF arising from the mixture of LOS/NLOS links.

### C. Angular Distribution

Having considered the path loss, we now turn to the path angles. Fig. 5 plots the distribution of the angles of the different paths in the links conditioned on the distance between the UAV and gNB. The conditional distribution is computed over the 10 strongest paths within each link for all links in the test dataset. For the sake of readability, we combine the aerial and terrestrial gNBs, and consider the total link distance (horizontal and elevation). Each row in Fig. 5 plots the conditional distribution of one of the four angles, $\phi_k^{rx}, \theta_k^{rx}, \phi_k^{tx}, \theta_k^{tx}$, relative to the LOS direction (even when an LOS path does not exist). The left-hand-side column is the conditional distribution of the angles for the test data. The right-hand-side column is the conditional distribution of randomly generated angles from the learned model.

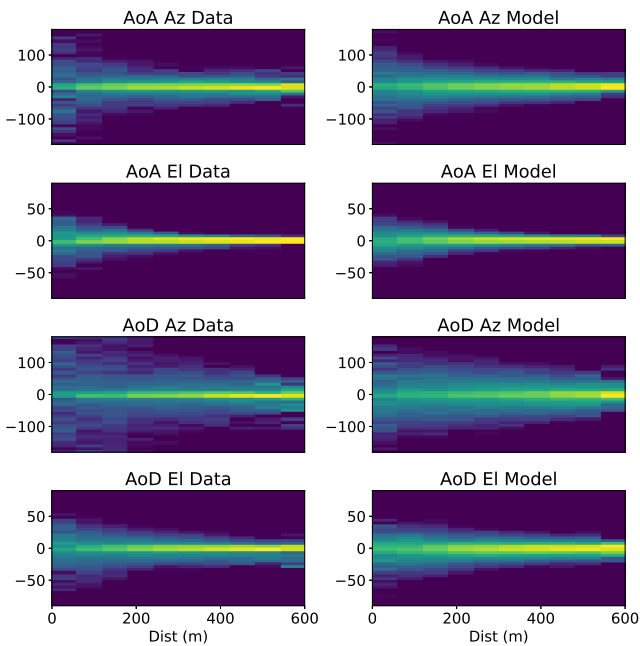

Fig. 5: Conditional distribution of the angles of the 10 strongest paths in each link relative to the LOS direction. Each row represents one of the four angles $\phi_k^{rx}, \theta_k^{rx}, \phi_k^{tx}, \theta_k^{tx}$. The left-hand-side column is the empirical condition distribution on the test data. The right-hand-side column is the distribution from the learned model.

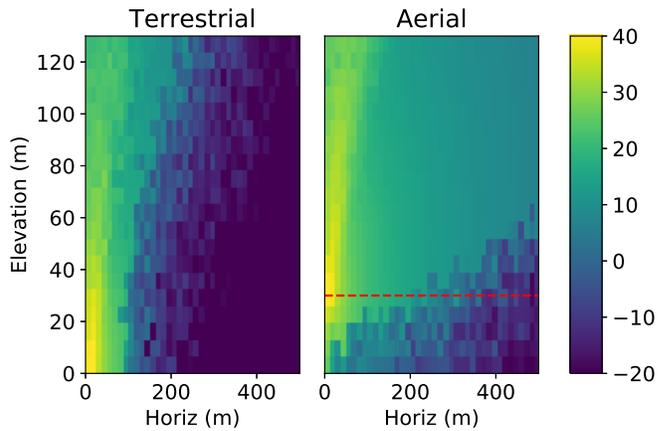

Fig. 6: Median SNR predicted by the model as a function of the horizontal and elevation position of the UAV. Details in Table II.

TABLE II: Uplink single-cell simulation parameters.

| Item | Value |
|---|---|
| Spectrum | Carrier frequency: 28 GHz |
| | Bandwidth: 400 MHz (4 × 100 MHz aggregation) |
| gNB height | Terrestrial: 2 m; Aerial: 30 m |
| Array size | UAV: $N_{\text{UAV}}$ = 16 (4 × 4 UPA) |
| | gNB: $N_{\text{gNB}}$ = 64 (8 × 8 UPA) |
| Array vertical angle | UAV: 180° ↓ lower hemisphere coverage [30] |
| | Terrestrial gNB: 100° ↘ ground coverage, 3 sectors |
| | Aerial gNB: 0° ↑ upper hemisphere coverage |
| Transmit power | UAV: 23 dBm |
| Losses | 6 dB including noise figures [31], [32] |

The model matches very well the general trends in the angular distribution. In particular, it captures an important phenomenon: for all distances and angles, the NLOS paths tend to be angularly close to the LOS direction. Moreover, the angular spread decreases as the UAV and gNB are further apart. This behavior makes intuitive sense in that, as the UAV pulls away from the gNB, there is less local scattering to create angular dispersion.

### D. SNR Predictions

We finalize with a demonstration of a simple application enabled by the generative model. Specifically, we compute the predicted uplink SNR as a function of the UAV position in the single-cell scenario described in Table II. A terrestrial or aerial gNB is located at $(0, 0, h)$ with $h = 2$ m and $h = 30$ m in the terrestrial and aerial cases, respectively. In the terrestrial case, the gNB is modeled as three-way sectored with half-power beamwidth of 90° per sector; the arrays in each sector have a 10° downtilt, as customary to serve ground users, hence the connections to UAVs must be through sidelobes or reflected paths [18], [33]. In the aerial case, the gNB is single-sectored with an upward-facing array intended for aerial coverage. The UAV, which features a single array at its bottom, designed for lower-hemisphere coverage [30], is at a position $(x, 0, z)$ with $x \in [0, 500]$ m and $z \in [0, 130]$ m. For each UAV position and gNB type (aerial or terrestrial), 100 channels realizations are generated by the model.

From the channel paths and the link budget values in Table II, which are consistent with current 28-GHz 5G deployments [30], the local-average wideband SNR is computed. Fig. 6 plots the median SNR, with the red dotted line indicating the antenna height of the aerial gNB.

The experiment shows how SNR predictions can be produced from the model and the array specifics. The aerial rooftop gNBs provide much greater coverage at large horizontal distances, yet terrestrial gNBs can provide very good coverage when the horizontal distance is small (less than 100m). This coverage from terrestrial gNBs is rather surprising: complying with the 3GPP model [1], terrestrial gNBs have downtilted antennas with a 30-dB front-to-back gain, precluding connectivity from direct vertical paths. However, the learned model captures local scattering from neighboring buildings within the antenna beamwidth, and the simulations show that these paths do enable coverage.

## VI. CONCLUSION

Generative NNs are a fitting engine for statistical channel modeling in complex settings. Provided that abundant data is available, they are perfectly equipped to learn intricate probabilistic relationships and then produce parameters distributed accordingly. The only assumption is the choice of the parameters themselves, which can rest on basic principles of radio propagation.

This paper has validated the methodology for an air-to-ground channel, in itself a prime example of complex setting, and specifically for an urban environment at mmWave frequencies. The resulting model, publicly available, has been shown to learn effectively and to make interesting and nonobvious predictions.